\documentstyle[aps,aps12,epsf,pre,floats,twoside,times,tighten]{revtex}

\begin{document}

\title{Actin, a model semi-flexible polymer }

\author{A. C. Maggs}

\address{PCT, ESPCI, 10 rue Vauquelin, 75231 Paris Cedex 05, France. tony@turner.pct.espci.fr}

\maketitle
{\begin{center} \bf For the Colloquium, Jacques Bastide \end{center}}

\begin{abstract}
 Actin is a filamentary protein which has many remarkable properties
 making it an ideal system for the study of the dynamics and
 mechanics of semi-flexible polymer solutions and gels; actin has a
 persistence length of over 10 microns and can polymerize to lengths
 of several tens of microns, this permits the use of video microscopy
 and other optical methods in the study polymer dynamics. Many
 associated proteins exist which can be used to regulate length or
 crosslink filaments. We discuss the dynamics and rheology of these
 solutions.
\end{abstract}

\subsection*{Actin, a self assembling protein}

For the rheologist the biological world is a source of a number of
interesting materials with properties which are not easily reproduced
with synthetic polymers. In particular actin forms remarkably stiff
structures which are very good realizations of semi-flexible
polymers\cite{hearst}. A number of experimental and theoretical
results are available for this system which we now review. The actin
experimental system is not without disadvantages, in particular it is
mechanically fragile and subject to bio-degradation, however it allows
the direct, visual observation of a number of fundamental processes in
polymer dynamics, including reptation. A number of novel regimes,
without direct equivalents in the case of flexible polymers, are still
to be explored.

\begin{figure}
  \epsfxsize=150pt \centerline{\epsfbox{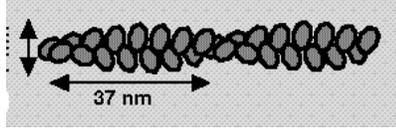}}
\caption[filament]{\label{snap}
  A schematic drawing of an actin filament showing the double helical
  structure. The two filaments are oriented in the same sense so that
  the global symmetry of the filament is polar as well as chiral.
  This should be contrasted with other helical biopolymers such as DNA
  where the two backbones are anti-parallel and thus form a non-polar
  structure. }
\end{figure}

Actin is a widely spread protein, occurring in almost all higher
cells. Starting from a rich source of actin (such as muscle) efficient
purification protocols are easy to apply, giving a relatively pure
solution of monomeric actin proteins. In the presence of a {\it
  polymerization buffer } \cite{purity} (containing divalent ions and
a source of energy in the form of the molecule ATP), the actin
monomers spontaneously assemble to form filaments, fig.  (\ref{snap}).
The filaments can also assemble without the energy source by replacing
the molecule ATP by ADP, however higher concentrations of actin
monomers are then required and the quality of the filaments formed
seems to be not as good. These filaments are made of double helixes of
monomers where each monomer is in an equivalent position. The diameter
of the filaments thus formed is about 7 nm. The filament length can
reach some tens of microns, a length scale easily studied with optical
techniques and at the limit between the microscopic and the
macroscopic, where micro-manipulation techniques using optical and
magnetic tweezers can be used to study macromolecular systems
undergoing Brownian motion.

A single filament of actin is just a little bit too fine to be seen
with direct optical microscopy, however addition of a fluorescent
probe renders the filament easily visible. These fluorescence
techniques have been used to determine the persistence length of the
filaments via {\it Brownian spectroscopy}. One records a large number
of images of a filament using a video camera and performs an analysis
in terms of normal modes. A simple theory of bending elasticity (see
below) gives the elastic constants and thus the persistence length
\cite{gittes,ott}. This persistence length is close to 15 microns,
with perhaps a moderate dependency on buffer conditions
\cite{hervecarlier}. The accuracy of these methods is very good with
error bars on the persistence length of less than 10 per cent. Thus
actin filaments are rather good realizations of semi-flexible polymers
with $d \ll l_p < L$, where $d$ is the filament diameter, $l_p$ the
persistence length and L the filament length. One should note that the
persistence length is enormous even compared with that of other
biological molecules. For comparison DNA has a persistence length of
30nm a factor of more than 100 smaller. Another cytoskeletal protein,
tubulin, forms structures which are even more rigid, and polymerize to
a similar length; microtubules thus behave as rigid elastic rods,
thermal fluctuations are probably to small to be important in most
situations.

The process which polymerizes the filaments is reversible (though
dissipative due to the turnover in ATP which occurs during the
assembly process). A given filament continually gains and looses
monomers at its ends. Due to the dynamic processes of filament
formation it is difficult to dilute filaments to study individual
filament dynamics in the absence of steric interactions; on extreme
dilution the solution depolymerizes. This depolymerization can be
prevented by using an associated molecule such as phalloidin which
modifies the disassembly dynamics or by working at a sufficiently high
concentration\cite{carlier}. Polymerization is usually carried out in
conditions such that the final distance between the filaments formed
is between $0.2\mu$ and $1\mu$. In these conditions the solution is
semi-dilute but with properties which must be contrasted with those in
semi-dilute solutions of normal flexible polymers . Firstly the volume
fraction is extremely small, of the order of 1 per thousand. Secondly
the sequence of length scales that we are used to in flexible polymers
has been changed. In the physics of flexible polymer solutions we are
used to the length hierarchy $d < l_p \ll \xi$, where $\xi$ is the
length scale of density fluctuations. Here we are in the regime, $d
\ll \xi \ll l_p$ with $\xi$ the typical distance between filaments
(d=7nm, $\xi=1\mu$, $l_p=15\mu$). We have already noted that it is not
possible to decrease the density to very low levels because of
filament depolymerization. Increasing the volume fraction leads to a
nematic transition\cite{nematic}. One is thus confined to a relatively
narrow window of concentration if one wishes to study isotropic actin
solutions.

As well as being studied by video methods, actin has been the subject
of investigation by light scattering, macroscopic rheology, magnetic
bead rheology and other physical techniques which we shall also
discuss. We note that this article, presented at the colloquium in
honor of the late Jacques Bastide, is partly review and partly a
summary of some unpublished ideas; some parts of the presentation have
been worked out in full, while others (such as the arguments as to the
nature of long range order in actin solutions) await disproof by more
rigorous approaches to the problem.

\subsection*{The length of actin filaments can be regulated by many proteins}

Actin is remarkable for the number of other proteins which interact
with it. Many tens of other proteins are available to modify the
properties of an individual filament or to regulate the degree of
polymerization. Some of these molecules have been exploited in
physical measurements but many other proteins are relatively poorly
studied. A general review of associated proteins can be found in
\cite{molecules}. These associated molecules are used to regulate the
rheological state of actin in the cellular environment.

Too many classes of regulating proteins exist to be discussed here,
but we will speak of some of the molecules which have the most
interest for the polymerist. Often used in rheological experiments,
gelsolin caps the end of a growing filament and thus can be used to
regulate the average filament length in a solution. When added to an
already polymerized solution gelsolin cuts the actin filaments in
presence of calcium, thus provoking a gel-sol transition. Since
unbound gelsolin also acts as a nucleating center for actin filament
growth one could imagine a protocol involving gelsolin and phalloidin
to generate relatively mono-disperse actin solutions \cite{protocol}.
It would be interesting to perform rheological experiments on such
mono-disperse samples to see how the usual polydispersity affects the
mechanical properties.

There are many other regulating proteins that interact with monomeric
actin, including profilin and thymosin-$\beta$4. These proteins
sequester actin in a non-polymerizable form and allow the cell (in
particular platelets) to stock very high concentrations of actin. A
chemical signal \cite{carlierprof} then starts an explosive
polymerization of filaments. This polymerization is into a disordered
phase but the concentration is high enough for nematic effects to be
important. There are some interesting problems in the rearrangement
dynamics to be answered. Does the solution re-orient via classical
rotational dynamics or does the dynamic growth and depolymerization of
the filaments play and important role in the re-orientational
dynamics? Or does the system remain blocked in a metastable glass like
phase as described by \cite{edwards}.

\subsection*{Observation of individual filament dynamics}

Despite the fact that most of the the solutions studied are
semi-dilute a number of the experimental results can be understood
from the dynamics of individual filaments, in particular recent
experiments of dynamic light scattering and micro-rheology with
magnetic beads. The dynamics of a free actin filament are determined,
for wavelength small compared with $l_p$, by the bending elastic
modes. To a good approximation,\cite{landau}, one can write that the
bending energy is
\begin{equation}
E={\kappa \over 2} \int {1 \over {R^2(s)}}\, ds \approx {\kappa \over 2} 
\int ({{d^2{\bf r_\perp}}\over{d s^2}})^2\, ds
={\kappa \over 2} \int q^4 {\bf r}_q {\bf r}_{-q} dq
\label{energy}
\end{equation}
Where $R(s)$ is the curvature of the filament at the position $s$,
$\kappa=l_p k_B T$ and ${\bf r}_{\perp}$ is the fluctuation of the
filament about its mean position. Eq. (\ref{energy}) shows that the
fluctuations in position are characterized by a $q^4$ dispersion
relation. Fitting the observed mode amplitudes to the functional form
(\ref{energy}) gives the bending constant $\kappa$ \cite{gittes,ott},
as noted above.

If we neglect hydrodynamic interactions (which give small corrections)
we find that the bending dynamics are described by a Langevin equation
\begin{equation}
 \rho{{\partial {\bf r}_\perp} \over {\partial t}}= -\kappa { {\partial ^4 {\bf r_{\perp}} \over { \partial s^4 }}} +{\bf f_{\perp}}(s,t)
\label{langevin}
\end{equation}
where $\rho$ is a friction coefficient and ${\bf f_{\perp}}(s,t)$ is
the force acting perpendicular to the filament. From eq.
(\ref{langevin}) we deduce the relaxation time for bending excitations
on the filament. Typical times for a ten micron filament are several
tens of seconds. Thus normal video cameras (with 30 frames per second)
are adequate for the study of filament dynamics. Higher resolution
methods (in time and space) involve interferometric and other optical
techniques\cite{weitz}, \cite{schmidt2}.

In the presence of a constant force, $ \delta(s) {\bf f_0}$, on the
filament at the origin one can solve the differential equation
(\ref{langevin}) for the displacement of the filament. To understand,
simply, the result of the full calculation one can make the following
simple scaling argument. The equation (\ref{langevin}) resembles that
of normal diffusion except that the second spatial derivative has been
replaced by a fourth spatial derivative. This implies that the
perturbation due to a small perturbation at $t=0,s=0$ will propagate
with the law $ { \bf r_{\perp} } \sim G( s^4/\kappa t) /(\kappa
t^{1/4})$. Thus after a time $t$ a length $(t \kappa)^{1/4}$ has been
perturbed and is in movement. The friction coefficient of this section
will increase with its length and thus with the same power law. The
approximate differential equation for the movement is thus $d {\bf
  r_{\perp} } /dt \sim {\bf f} / (t \kappa)^{1/4} $. Integrating this
equation we find that under constant force, ${\bf f_0 }$,
\begin{equation}
{\bf r}_{\perp}(0,t) \sim 
{{\bf f_0} t^{3/4} \over {\kappa^{1/4}}}
\label{force}
\end{equation}

\begin{figure}
  \epsfxsize=200pt \centerline{\epsfbox{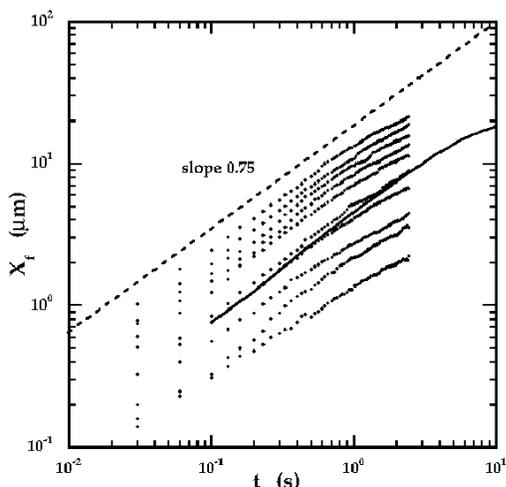}}
\caption[Amblard]{\label{slope} 
  Displacement as a function of time of a bead in a actin network
  under a constant force. The scale is log-log. The dashed line
  corresponds to a slope of 3/4 as predicted by eq. (\ref{force}).
  Each curve corresponds to a different applied force, allowing one to
  check the linearity of the response as a function of ${\bf f_0}$.
  For long times and larger forces there is a saturation of the
  response, presumably due to steric hindrance coming from
  interactions between several filaments.}

\end{figure}

Such experiments have been realized using small magnetic beads
\cite{amblard,ziemann,zaner}, fig. (\ref{slope}). One polymerizes a
solution of actin in presence of a small number of magnetic particles.
One chooses the size of the beads such that they are trapped by the
mesh formed by the actin filaments but the beads are small enough that
one is sensitive to individual filament fluctuations. Thus the
particles are trapped in a small cage and pulling on the bead disturbs
only a few filaments. The experiments are in very good agreement with
the prediction (\ref{force}), both the power law dependence in time
and the linearity in force. Independent examination of the
experimental curves of other experimental groups is also consistent
with the anomalous time dependence in eq. (\ref{force})
\cite{ziemann}. The forces applied in these experiments are very large
compared with the thermal energy scale: they are of the order of $1
pN$, whereas the {\it natural}\/ force unit for this system would be
$k_BT/\xi$ which is much smaller, comparable to $10^{-3} pN$. These
experiments are thus only weakly sensitive to thermal fluctuations;
the force constraining a given filament to its tube is also comparable
to $k_BT/\xi$ perhaps explaining why tube constraints do not seem to
be important in these experiments, and why one is able to displace the
bead rather large distances before eq. (\ref{force}) breaks down.

Many approximations are involved in the application of (\ref{force})
to actin solutions. On further reflection is is even surprising that
the experimental data show such good power law behavior over several
decades. Some of the points to consider are

\begin{itemize}
 
\item The approximation of a point force acting on a filament: The
  beads must couple by hydrodynamic interactions to a finite portion
  of the filament. Thus one expects to see important corrections to
  (\ref{force}) due to crossovers in hydrodynamic couplings. In
  addition, at the shortest times the friction of the bead should be
  larger than of the section of filament of length $(t \kappa)^{1/4}$,
  due to the constant friction on the bead At longer times constraints
  due to filament contact must also become important.
 
\item Eq. (\ref{force}) is only valid in the absence of tension.
  Addition of a term in $\sigma \partial^2 {\bf r_{\perp}}/ds^2$ in
  eq. (\ref{langevin}) should change the exponent in (\ref{force}) to
  $1/2$. Such tension terms could be generated by crosslinks, or
  disruption of the equilibrium structure of the network under large
  forces
 
\item The bead can be dragged through the sample in a stochastic
  manner as pores in the network open and close \cite{ziemann}. We
  also note that a very interesting non-linear dynamic regime has been
  discussed \cite{nelson}, where thermal fluctuations in a filament
  are pulled out via a large forces. This too may be important when
  dragging beads through the sample.

\end{itemize}

The nature of the crossover from local dynamics dominated by bending
elasticity described by (\ref{force}) to the macroscopic bulk modulus
has yet to be full explored. It is not clear whether such a crossover
should occur for particles of size comparable to $\xi$, or whether
there is a novel viscoelastic regime all the way out to $l_p$.  Recent
experiments \cite{schmidt2} observing thermal fluctuations of beads
with sizes up to 20 times the mesh (but smaller than $l_p$) still seem
to be sensitive to the internal mode dynamics, though they are
interpreted in terms of macroscopic elastic theory. We argue later
that elasticity on the scale $\xi$ to $l_p$ is non-classical, perhaps
resembling smectics or columnar liquid crystals rather than isotropic
solids.

Individual filament dynamics can also be studied in dynamic
quasi-elastic light scattering experiments. One illuminates a solution
with a coherent source which is scattered by a wave-vector $q$ and
observes the time-time correlations in the scattered beam.  The Munich
biophysics group \cite{schmidt,piek} observed unusual results in their
light scattering experiments. In particular they found that the
results scaled with the combination $tq^{2.7}$ rather than $tq^3$ or
$tq^4$ expected in Rouse or Zimm dynamics \cite{doi,degennes}. The
dynamic light scattering experiments measure the density density
correlation function. These experiments are again sensitive to the
transverse fluctuations of the filaments described by eq.
(\ref{langevin}) with ${\bf f_{\perp}}(s,t)$ now interpreted as the
Brownian noise of the solvent. The transverse fluctuations can, as
usual, be characterized by the correlation function
\begin{equation}
G(s,t) = < ({\bf r_{\perp}}(s,t)-{\bf r_{\perp}}(0,0))^2 >
\label{G}
\end{equation}
indeed the experiments are sensitive to
\begin{equation}
I(t,q) = \int \exp( -q^2 G(s,t)/4)\, ds
\end{equation}
For times that are not too short one finds that the correlation
function is dominated by the contribution coming from $s=0$ and that
we need only calculate \cite{farge}
\begin{equation}
G(s=0,t) = \int { {1- \exp(-2 \kappa q^4 t/\rho)} \over {\kappa q^4} } {{dq}\over{2 \pi} }= A t^{3/4}/\kappa^{1/4}
\label{GG}
\end{equation}
thus
\begin{equation}
I(t,q) \sim \exp( -A' q^2 t^{3/4} /\kappa^{1/4})
\end{equation}

The intensity scales with the combination $tq^{8/3} \approx tq^{2.7}$
as found in the experiments. More detailed calculations, including
hydrodynamic interactions, allow one to extract a value for $\kappa$
from the light scattering data which is consistent with that found in
Brownian spectroscopy \cite{farge},\cite{kroy}. Recently a more
detailed study has been performed to examine the short time behavior
\cite{kroy,sackmannkroy} where the full functional form of (\ref{G})
has to be examined. Other experimental studies have confirmed these
results\cite{archen}. Another dynamic light scattering approach to
measuring the rigidity of actin filaments involves using short
filaments and studying corrections to the rigid rod
behavior\cite{archen} in an expansion in normal modes of the filament.

\subsection*{Tube dynamics of actin filaments}

A rather pretty series of experiments have been performed \cite{kas}
in order to visualize the movement of actin filaments in semi-dilute
solutions. As we have already remarked it is possible to examine a
filament with an optical microscope by adding a fluorescent probe to
the sample. However, marking an entire semi-dilute sample swamps any
image detection system. One gets around this problem by using a two
stage polymerization process. Firstly polymerizing in presence of a
dye, fixed to phalloidin, secondly diluting the fluorescent actin in a
new polymerizing solution containing unmarked phalloidin. In this way
non-fluorescent filaments reform around the now dilute marked
filaments. Thus one directly observes filament movement, the existence
of the tube, as well as events such as constraint release as the
geometry of the tube changes in time.

Equally important for the general study of actin solutions the
observations show that the samples studied are not too contaminated by
crosslinking and cutting agents (see below) and that the samples
studied are really semi-dilute fluids rather than gels: A major
problem in the literature on actin mechanics is that it is very
difficult to have precise information on the degree of crosslinking,
or the distribution of filament lengths in a given sample. This leads
to many divergences between experimental groups as to the state of the
sample. Joint mechanical/optical experiments such as those performed
already in \cite{cortese} should lead to a much better understanding
of the problems of sample preparation.

While the dynamics of a free filament is largely dominated by
transverse fluctuations these modes are clearly blocked by the tube
constraints \cite{degennes,doi}. Thus one expects that there is a
characteristic length $l_e$ below which the filament fluctuations are
essentially free but beyond which fluctuations are suppressed. This
length is comparable to the distance between the filaments, $\xi$, but
does show a slightly different scaling $ l_e \sim \xi^{4/5} l_p^{1/5}
$, \cite{herve,semenov,semenovbis}. As a consequence of the presence
of several length scales in actin solutions direct application of
scaling ideas in dynamics is sometimes a difficult and ambiguous
enterprise.

\begin{figure}
  \epsfxsize=200pt \centerline{\epsfbox{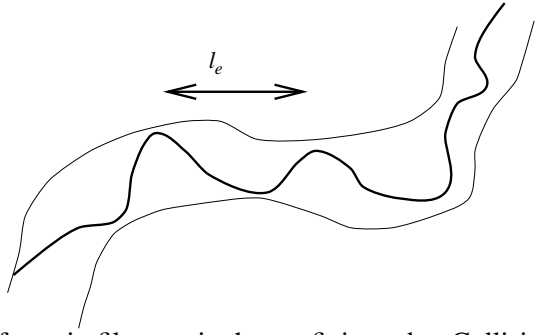}}
\caption[tubecap]{\label{tube}
  
  The fluctuations of a actin filament in the confining tube.
  Collisions occur between the tube and the filament every $l_e$.
  Notice that the filament is rigid on the scale of $l_e$ which should
  be contrasted with the case of flexible polymers where one imagines
  a Gaussian, or Flory like structure on length scales smaller that
  the tube diameter.  }
\end{figure}

In the interior of the tube the filament behaves in a very anisotropic
manner. A segment of filament of length $l_e$ is characterized by two
elastic moduli, The first characterizes its transverse response and
fluctuations and is quite soft. The second which characterizes the
longitudinal fluctuations and response is much harder. To understand
the macroscopic elastic behavior of a actin solution we need to
understand these two moduli and their coupling to macroscopic degrees
of freedom. The transverse elastic constant of a filament of length
$l_e$ follows from eq. (\ref{energy}). Applying a static force ${\bf
  f}_{\perp}$ transverse to a filament gives rise to a displacement
${\bf r}_{\perp} \sim {\bf f}_{\perp} l_e^3/l_p$, thus we characterize
the section of filament by a transverse elastic constant
\begin{equation}
K_t \sim k_B T l_p/l_e^3
\label{Kt}.
\end{equation}
This elastic response is purely mechanical in origin.

The origin of the longitudinal elastic constant is however more
subtle; thermal fluctuations ``eat up'' a part of the length of a
filament. Imagine a filament aligned at temperature zero along the z
axis. The addition of thermal fluctuations in the x-y plane will cause
the filament to shorten in the z direction. Thus the projected length
will fluctuate.  The extra length of a filament undergoing small
fluctuations about its mean position can be calculated from
\begin{equation}
S= \int_0^{l_e} (\sqrt{ 1+ (d{\bf r_{\perp} } /ds)^2 }-1)\, ds 
\approx 1/2 \int_0^{l_e} (d{\bf r_{\perp} } /ds)^2 \,ds
\label{length}
\end{equation}
Using equipartition for the energy (\ref{energy}) one finds that the
mean excess of material in a tube of length $l_e$ is
\begin{equation}
S \approx l_e \int_{1/d}^{1/l_e} {q^2\over{\kappa q^4}} \sim l_e^2/l_p
\label{length2}
\end{equation}
Similarly the elasticity is calculated by examining the fluctuations
of $S$ in particular ($\langle S^2 \rangle -\langle S \rangle ^2$). A
similar calculation to (\ref{length2}), \cite{herve,fred,fred2} shows
that this fluctuation can be interpreted as an elastic constant
\begin{equation}
K_l \sim k_B T l_p^2/l_e^4
\label{longitudinal}
\end{equation} 
When $l_e \ll l_p$ the longitudinal constant becomes very hard and
$K_l \gg K_t$. According to whether one is coupled to the transverse
or the longitudinal fluctuations one feels very different elastic
constants.  In the case of micro-rheology, discussed above one is
sensitive to the transverse fluctuations, in the case of macroscopic
rheological measurements to be discussed below the larger longitudinal
elastic constant can dominate.

\subsection*{Elasticity of crosslinked actin}

In \cite{fred} the authors consider the effect of a large degree of
crosslinking on actin solutions and conclude that in this case the
longitudinal constant dominates the macroscopic rheological behavior.
They find that the shear modulus, $G \sim k_B T l_p^2/l_e^5$. However,
as a function of the degree of crosslinking one may expect a number of
distinct regimes. The modulus is high at high crosslinking densities
because the constraints on the distances become very severe and it is
impossible to shear the sample without compressing longitudinally the
filaments. At low degrees of crosslinking this is no longer the case;
one can imagine configurations where the sample is linked in a three
dimensional manner but the number of constraints is to low to impose
compressive strains on the actin filaments.  \footnote{An analogy is
  the problem of regular lattices in two dimensions: A three
  dimensional triangular lattice is rigid due to the fact that the
  number constraints is superior to the number of translational
  degrees of freedom. The hexagonal lattice however is floppy since
  the number of constraints per lattice site is lower. The marginal
  case is the square lattice. See \cite{aside}} In such loosely
crosslinked systems, under shear, there is no reason either to bend or
to compress the filaments (unless the crosslinks themselves impose
specific angles between the filaments).  In cases of weak crosslinking
one is probably dominated by entropic contributions to the free energy
and should count $k_BT$ towards the modulus for each effective degree
of freedom in the gel.  This would lead to a low modulus comparable to
$k_B T/\xi^3$. If the filaments are crosslinked by angularly stiff
elements which impose a given angle between two filaments one can only
shear the sample by bending the individual filaments; in such a case
one would estimate $G \sim l_p/l_e^2\xi^2$, due to a coupling to
$K_t$. Thus, for crosslinked samples one expects a wide variety of
elastic moduli as a function of the type and concentration of
crosslinking elements in the solution.

There are clearly some interesting problems in the percolation of
elastic constraints \cite{percolations} in actin networks. Indeed the
nature of the transmission of stress by directional and
non-directional bonds in diluted networks has been intensively studied
over the last decade and one finds that there are very different
critical behaviors as a function of crosslinking. In particular the
percolation threshold for bending and compressional degrees of freedom
can be rather different, as is also the case for the exponents. There
must surely be a rather rich series of crossovers in the presence of
actin crosslinking agents and as a function of mean filament length
due to the large difference in $K_t$ and $K_l$. The exact nature of
the possible regimes still have to be elucidated however. In fact the
nature of the percolation transition in even simple elastic networks
is still an active subject of research \cite{dux}.

We note also that the nature of entanglements is very different in
semi-flexible solutions compared to solutions of flexible polymers.
Because of the convoluted nature of the path followed by flexible
polymers in space knots, or entanglements, are very easily formed on a
length scale which is comparable to $\xi$. In the case of actin
filaments there are important transverse constraints on the scale of
$l_e$, which is comparable to $\xi$, however there is no longitudinal
hindrance to movement. We shall argue that despite the lack of
parallel entanglement the storage modulus of actin solutions can still
be rather high.

\subsection*{Shear couples to length fluctuations}

\begin{figure}
  \epsfxsize=260pt \centerline{\epsfbox{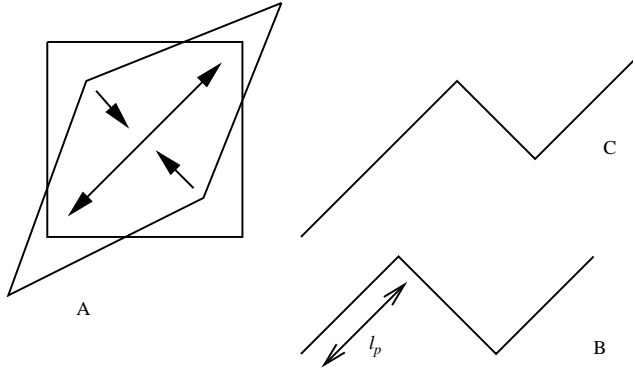}}
\caption[shear]{
\label{shear}
Coupling of shear to length: (A) The shear has both rotational and
elongational components, so that a tube with initial shape (B)
containing a filament is alternatively lengthened and shortened (C). }
\end{figure}

One of the most important differences in the mechanics of
semi-flexible and flexible polymers concerns the coupling of external
shear to length fluctuations of the molecule in a semi-dilute
solution. Shear does not couple to the length fluctuations of a
flexible polymer in linear response \cite{doi}. This is due to the
fact that the confining tube contains a coil in which all orientations
are present; the extensional component of the shear is thus unable to
couple to the orientation of the polymer tube. The situation is very
different for actin filaments where no averaging occurs within the
tube. One expects that locally the filament can be placed under
tension or compression according to its orientation with respect to
the extensional component of the shear fig. (\ref{shear}).

The exact coupling of actin filaments to a shear flow is difficult to
calculate. We have calculated \cite{me} the contribution of the
longitudinal degrees of freedom to the modulus using a single chain in
mean field picture with the usual assumptions about affine shear in
the sample.  We estimated the modulus \cite{me} by calculating the
relaxation of the stress in a sample to which a sudden shear,
$\gamma_0$, has been applied. When we apply a macroscopic shear to a
actin solution a test chain will feel the effect of the distortion via
hydrodynamic coupling with the tube.  During the fast application of
the shear the internal modes of the filament relax quickly.  For the
later relaxation of the sample we have the following relationship
between the stress $\sigma(t)$ and the modulus $G(t)$, \cite{doi}
\begin{equation}
\sigma(t) = \gamma_0 G(t)
\label{shearq}
\end{equation}
Thus we calculate the modulus \cite{me} by studying the relaxation
from an initial perturbed configuration of alternate compression and
tension, fig. (\ref{shear}). We sumarise the results in the next two
sections.


\subsection*{Time scales in semi-dilute solutions}

A number of interesting times scales exist in semi-flexible polymer
solutions which have no direct analogy in the dynamics of flexible
polymers. Indeed the nature of the time scales and dissipation
occurring in semi-flexible polymer solutions is sufficiently different
from those in flexible polymers that the experimental similarities
between the two must be considered rather fortuitous. The shortest
characteristic bending time in our problem is that corresponding to
the length scale $l_e$. From eq. (\ref{langevin}) one can estimate
that the tube constraint becomes important for $\tau_e \sim l_e^4/l_p$
\footnote{Note that we choose for the rest of this
 article to work in the {\it natural}\/ units for this problem; all
 lengths are in microns while time has the units of microns cubed.
 This is rather convenient because the volume 1 $\mu^3$ corresponds
 to a few seconds. }.
 Thus $\tau_e$ is typically somewhat smaller than a
second. The longest natural time in the problem is that needed for a
filament to diffuse completely out of its tube. From the diffusion
equation $ R^2= D t$ and using the fact that the diffusion coefficient
varies with the length of a filament as $D_L \sim 1/L$ one finds that
the reptation time, $\tau_{rept} \sim L^3$. For filaments which are
tens of microns long this latter time can be hours, as seems to be
seen in experiments of the Munich group \cite{ruddies,muller}.
However, we note that even a small degree of crosslinking (far below
that needed for gelling) will we sufficient to produce star like
clusters for which the dependence of the relaxation time on the length
is exponential, $ \tau_{rept} \sim \exp(\alpha L)$, probably beyond
any experimentally reasonable limit.

Between these two times a number of other important time scales exist
which we shall now discuss. These other time scales come from the
relaxation of the internal longitudinal modes of the filament. As we
have argued above, for length scales longer than $l_e$ the degrees of
freedom available to a filament are longitudinal. The filament can
thus be modeled as a series of hard springs, of elasticity $K_l$. Each
spring has a relaxation time of $\tau_e$. The dynamics of each cell
are coupled together, the long wavelength dynamics are those of a one
dimensional elastic chain in a dissipative environment. Thus one can
write
\begin{equation}
 { {\partial r} \over {dt}} = D_e {{\partial^2r} \over { \partial s^2}}
\label{diffusion}
\end{equation}
where $D_e$ is the effective spring constant over a dissipation
coefficient, $r$ is the longitudinal displacement. Since we know that
the time $\tau_e$ corresponds to the length scale $l_e$ one expects
that $D_e \sim l_e^2/\tau_e $.

From eq. (\ref{diffusion}) one can deduce other characteristic time
scales. Firstly $\tau_{L}$ the relaxation time for internal modes
in the whole filament and secondly $\tau_{l_p}$ the relaxation time
for fluctuations coherent over the persistence length $l_p$. Since
eq. (\ref{diffusion}) has the form of a diffusion equation we have
that $\tau_{L} \sim \tau_e (L/l_e)^2$ and that
\begin{equation}
\tau_{l_p} = \tau_e (l_p/l_e)^2 
\label{relax}
\end{equation}
This final time scale is most important in the macroscopic relaxation
of density fluctuations. As we have seen above shearing a sample leads
to stresses which are coherent over the length $l_p$, they can only
relax slowly via eq. (\ref{diffusion}) on a time scale comparable to
$\tau_{l_p}$. This time is of the order of a tens of seconds under typical
experimental conditions. A number of experiments of the Munich group
\cite{ruddies,muller} indeed have a dynamic crossover at this time
scale in macroscopic rheological experiments.
A fuller discussion of time scales is given in \cite{herve}.

\subsection*{Plateau Moduli of actin solutions}

Much of the rheological literature on actin solutions has concentrated
on the existence and value of a plateau modulus. Analogies are often
made with the rheology of flexible polymers where an elastic plateau
is seen. To what extent can one take over the standard results of
polymer rheology to the case of semi-flexible polymers with the length
hierarchy $d \ll \xi \ll l_p$? A number of models have been proposed for
the plateau modulus of actin solutions. However many of these models are
probably too mechanical to be taken seriously. In particular to delimit
the frequency range in which a plateau is observed one needs some
ideas of the dynamical processes  and the mechanisms of
dissipation present as well as the dynamical mode implicated in energy
storage. Many of the simplest models in the literature have no
dynamic component \cite{biophys,freyprl}.

The quality of the plateau observed is sometimes not very good; for
instance we can cite the creep experiments of \cite{quarter}, or of
\cite{creep}. Other experimental groups see quite a flat plateau
\cite{janmey,janmey2}. It is unclear why the experimental literature
is so disperse. Perhaps a number of experiments are in fact seeing a
slow variation of the storage modulus rather than a real plateau.
Certainly the experiments are difficult to perform and there are
perhaps important and interesting effects coming from
polydispersity, nonlinearity and sample history which need to be
cleared up. We believe that two mechanisms are able to give rise to
real plateau regimes and that some of the discrepancies in the
experimental literature could be due to the fact that different
storage processes are being observed in different experimental setups.
However, we shall also argue that over certain frequency ranges, one
might expect a behaviour in $G(\omega) \sim \omega^{1/4}$ for some
systems or even a stretched exponential decay in $G(t)$. These slow
frequency variations could in some cases be confused with a poor
quality plateau.

We have argued \cite{herve} that on time scales larger than $\tau_L$
all longitudinal stress in a filament has been dissipated, but that
there is still a large frequency range between $\tau_L\sim L^2$ the
slowest internal mode and $\tau_{rept}\sim L^3$. Over this time scale
the backbone of the tube is in a sheared, non-Gaussian form, there
must be some residual entropic contributions to the free energy due to
this modified tube geometry. We have thus suggested a small
low-frequency contribution to the modulus of the form $G \sim
k_BT/\xi^2 l_e$ \cite{herve}. This comes from the modification of the
confinement energy in the tube which is equal to $k_BT$ for each
segment of length $l_e$ of filament in the solution. This very low
modulus seems to be seen by some experimental groups working with very
low frequency viscoelastic response \cite{ruddies,muller}. Recent
experiments interpreting bead fluctuations also find small elastic
moduli \cite{schmidt2}.

As we have noted many groups see a much larger modulus at higher
frequencies, how can one explain these results? As argued above the
effect of shear on a filament is to produce tension and compression on
filaments which are coherent over a distance of the order of $l_p$.
Since the longitudinal displacement variable, $r$ of eq.
(\ref{diffusion}) follows a diffusive type law the applied strains
remain high on a time scale varying from $\tau_e$ to $\tau_{l_p}$ (ie
fractions of a second to a few tens of seconds). One expects a see a plateau
over this frequency range which is high since one couples to the
longitudinal elastic constant $K_l$. In \cite{fred} a high modulus of this
form was derived by assuming that the solution is completely entangled
on the scale $l_e$ and that longitudinal dynamics as well as
transverse dynamics are hindered. We see that this hypothesis is
unneeded, the dynamics of longitudinal fluctuations are already very
slow due to the large length scales involved. This slow dynamics gives
a modulus similar to that estimated in \cite{fred} for the case of
strong local entanglement, $G\sim l_p^2/\xi^2l_e^3$.

At longer times there are two possibilities as a function of the
lengths of the filaments. For filaments which are only somewhat longer
than $l_p$ one expects that the stress in a filament should decay to
zero {\it exponentially}\/ for long times, when $t > L^2/D_e$. However
if the distribution of filament length is itself exponential \cite{dogterom}
a convolution of the length distribution with the exponential decay of
stress in each filament should give a stretched exponential form to
the function $G(t)\sim \exp(-(t/t_0)^{1/3})$.

In the case of very long filaments the following simple scaling
argument can be applied. The stress in a long filament is a random
alternation of compression and tension which we can consider as a
sequence of random positive and negative initial conditions in a
diffusion equation. Since the initial conditions in space are random
the Fourier coefficients of the initial condition are close to
Gaussian white noise. Each mode in the system decays as $a_q \sim
\exp(-D_e q^2 t)$. After a time $t$ only modes with $q< 1/\sqrt{D_et}$
remain. Again these modes are Gaussian, of mean zero and independent.
Going back into real space one thus concludes that the typical value
of the stress in a filament must decay as $ (D_et)^{-1/4} $ and thus
that the system as a contribution to the low frequency modulus of the
form $G(\omega) \sim (D_e^{-1} \omega) ^{1/4}$, as confirmed by more
elaborate treatments \cite{me}.

\begin{figure}
 \epsfxsize=250pt \centerline{\epsfbox{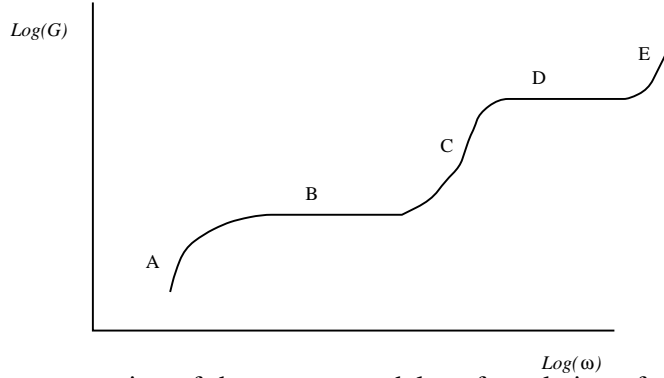}}
\caption[plateufig]{ \label{plateaufig}
 Schematic representation of the storage modulus of a solution of
 actin filaments showing a number of regimes (A)
 fluid behaviour at very low frequencies. (B) entropic plateau. (C ).
 crossover between the two plateau, which is expected to be a
 sensitive function of the distribution of filament lengths in the
 solution. (D) plateau due to longitudinal elasticity. (E) crossover
 to independent filaments.
}
\end{figure}

To resume, we believe two mechanisms are present, fig. (\ref{plateaufig}),
which can lead to plateau like behaviour in different frequency
ranges. No experimental data is available to confirm this point, but
the experiments from the various groups use a variety of experimental
methods and preparation conditions. The data is not always available
over the complete frequency and dynamic range that one might wish. The
crossovers between these two regimes can be quite rich and varied,
depending on the sample preparation.

\subsection*{Breakage, work-hardening and other non-linearities}


Solutions of actin filaments have a relatively small strain where one
observes linear response. There are a number of mechanisms which lead
to non-linear effects.

\begin{itemize}
\item The filaments are relatively fragile \cite{janmey,janmeykas}.
  Pipetting a sample breaks longer filaments, the act of filling a
  rheometer is thus relatively delicate. Best results are to be
  expected from in-situ polymerization.
 
\item Even very small shear rates are capable of aligning samples. In
  long experiments where the filaments are still dynamic one could
  worry about progressive alignment of filaments due to the flow
  \cite{cortese}, even in samples polymerized in situ.
 
\item Extensional elasticity described by eq (\ref{Kt}) depends on the
  small fluctuations in length of each cell of size $l_e$. From
  eq.~\ref{length2} we see that a filament has about 10 \% excess
  material. Strains larger than this exhaust the excess length
  generated by the fluctuations, the sample should thus become
  extremely non-linear for larger strains.
 
\item It is relatively easy to buckle filaments. Under oscillatory
  excitation one can imagine that filaments are able to buckle out of
  the confining tube forming {\it hernias}, similar to those observed
  in DNA electrophoresis. Once a hernia has formed it should
  considerably change the nature of the longitudinal tube dynamics.
  Multiple interacting hernias could even act as longitudinal
  entanglements and serve as a source of work hardening in actin
  solutions.
 
 
\item Very low levels of crosslinking, below that needed to produce
  gelling can introduce exponentially long, $\tau_{rept} \sim
  \exp(\alpha L)$, relaxation times in solutions. In creep experiments
  this will probably drive the system into highly non-linear regimes,
  since the slowest modes are very far from equilibrium.

\end{itemize}

\subsection*{Actin, Peirls and St Vernet }

What is the nature of the deformation field around a particle, such as
a small magnetic bead, on length scales between $\xi$ and $l_p$. We
argue here that it can be rather different from that found in
conventional elastic solids. Normal elastic materials are described by
longitudinal or transverse elastic modes, each solutions of Laplace's
equation. Thus the static deformations can be considered the solution
of the equation $\nabla^2 \phi = 0$ with the appropriated boundary
conditions; for normal solids we are interested in systems with a
$q^2$ dispersion relation. This has two important consequences
\begin{itemize}
\item As pointed out by Peirls there is long range order in three
  dimensions due to the infra-red convergence of the integral $\int
  d^dq/q^2$ in three dimensions (to be contrasted with the divergence
  in one and two dimensions).
\item As pointed out by St Vernet a force applied over a radius R will
  also decay in a material over a distance which scales as R.
\end{itemize}
We are interested in how these points are modified in actin solutions.
We first resume the Peirls argument in a real space form which can be
applied to the actin case with just a little modification. When we
apply a force on a solid then we can suppose that the solid reacts in
a sphere of diameter $L$. If the amplitude of the displacement at the
point of force is $a$ then the total energy stored is equal to $ E
\sim (a/L)^2 L^d$ in $d$ dimensions. The first term in the energy
comes from the bending elasticity described by the $q^2$ dispersion
law and the second factor is the volume excited. In three dimensions
this energy diverges with $L$ thus small forces only excite movements
locally about the point of force application. In low dimensional
systems we can disorder the system with the application of arbitrarily
small forces since $E$ goes to zero for large $L$.

Let us now pass to actin. If we consider only the transverse component
of the fluctuations then we have seen that actin solutions are
characterized, on lengths less than $l_p$ by a $q^4$ dispersion law.
Naive application of the Peirls arguments would then suggest that
there is no long range order on scales out to $l_p$, (We expect that
the system is described by an normal $q^2$ law beyond $l_p$). We now
give a more careful argument to show that the longitudinal and
transverse modes are probably strongly coupled so that long range
order is almost restored even for lengths smaller than $l_p$, however
the same argument gives an anomalous penetration of excitations into
the gel which should be contrasted with the result of St Vernet.
There is also an important difference in the scaling of the amplitude
of response with the particle size. In the case of standard elasticity
the amplitude of the response scales inversly with the particle size,
for actin the response is predicted to be independent of the size of
the probing beads.

The weak existence of long range order also has implications as to the
coherency of the picture of a tube in the dynamics. The tube geometry
may fluctuate more in semi-flexible polymer solutions than in
classical solutions of flexible polymers.

\begin{figure} 
  \epsfxsize=250pt \centerline{\epsfbox{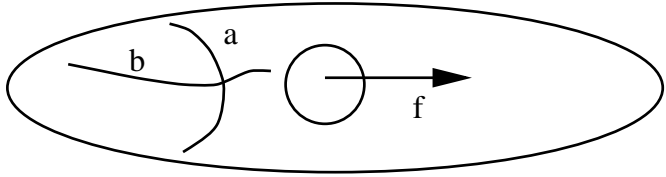}}
\caption[vernet] {\label{vernet}
 
  Pulling on a bead in a actin solution; bead is pulled in the
  direction marked $f$. The effect of the movement is confined to an
  ellipsoidal volume, length $L_l$ and width $L_w$. Filaments which
  are perpendicular to the force (such as filament ``a'') bend in
  response to the external force. Those filaments which are almost
  parallel to the applied force (filament ``b'') are under
  longitudinal stress. The total energy is thus the sum of two
  contributions as discussed in the main text. }

\end{figure}

Consider an excitation in an actin gel (perhaps excited by a small
bead on which one pulls with a constant force) fig.~(\ref{vernet}),
characterized by its width $L_w$ and its length (parallel to the
force) $L_l$. The energy of this excitation has two sources, bending
of filaments perpendicular to $f$ and longitudinal compression of
filaments parallel to $f$. The first contribution can be estimated as
\begin{equation}
E_1 \sim  (L_w^2 L_l) (k_BT l_p a^2/L_w^4) (1/\xi^2)
\label{e1}
\end{equation}
with the first factor is the excitation volume, the second factor the
energy eq. (\ref{energy}) and the third factor the length of filament
per unit volume. The longitudinal contribution to the energy is given
by
\begin{equation}
E_2 \sim  (L_w/\xi)^2 (l_e/L_l) (a^2 k_BT l_p^2/l_e^4)
\label{e2}
\end{equation}
The three factors here are the number of filaments crossing an area
$L_w^2$, the number of segments of length $l_e$ in the length $L_l$,
and the longitudinal spring constant $K_l$. If the bead sets the scale
for the transverse variations then minimizing the sum of the energies
over $L_l$ implies that
\begin{equation}
 L_l \sim L_w^2
\label{nonvernet}
\end{equation} in contrast with classical elastic behaviour. 

Note, in this suggestive argument, there are a number of hidden
assumptions as to the time scales involved. For the description of the
longitudinal modes in terms of longitudinal elasticity to be valid we
must be studying times such that $t > \tau_e$. Clearly for the
description in terms of a solid rather than a liquid to be valid we
require $ t < \tau_{rept} $. However the longitudinal stress can only
be maintained for the shorter time span of $t < \tau_{l_p}$. Thus we
might hope that the above description is valid for
$\tau_e<t<\tau_{l_p}$.

In three dimensions the total energy $E_1+E_2$ (obtained by
substituting (\ref{nonvernet}) in (\ref{e1}) and (\ref{e2})) remains
bounded, suggesting that the lower critical dimension for this problem
is three dimensions and that the amplitude of fluctuations diverges
logarithmically with distance up to a scale of $l_p$, rather like
elastic solids in two dimensions, or certain liquid crystals in three
dimensions. It will be interesting to try and understand how this
argument generalizes to times shorter than $\tau_e$ to fully
understand the experiments \cite{fred2}.

\subsection*{Statics and dynamics of twist}
  
Until now we have been concerned largely by the bending dynamics of
actin filaments, however an elastic filament has two degrees of
freedom. It is characterized both by bending and torsional elastic
constants. These two quantities should be comparable; for a cylinder
with Young modulus $E$, Poisson ratio $\sigma$, and diameter $d$ we
can calculate that the bending elastic constant and twist elastic
constants are given by \cite{landau},
\begin{eqnarray}
\kappa_{twist} &=& E \pi d^4/64(1+\sigma) \\
\kappa_{bend} &=& E \pi d^4/64 
\label{bending}
\end{eqnarray}
Since the Poisson ratio is normally between 0 and 1/2. the torsional
persistence length $l_r$ must be comparable to $l_p$. Clearly the
approximation of a uniform cylindrical material is not very good, and
one is sensitive to the internal modes of the filament which may give
important corrections \cite{computer}.

Most experiments discussed up to now have only been sensitive to the
bending degrees of freedom. Firstly because they couple most directly
to the bending modes. Secondly because the time scales of bend and
twist dynamics are very different as we shall now discuss. One of the
few experiments which is directly sensitive to the twist dynamics of
actin filaments is \cite{twist} where the dynamics are observed by
depolarized light scattering. With this technique the sample is marked
with a fluorescent dye and illuminated by polarized light.  During the
time between the absorption and the emission of photons the filament
can turn and thus the distribution of emitted light depends on the
twist dynamics. In principle this should allow a determination of the
twist persistence length. The experiments of \cite{twist} find rather
strange behavior of the measured constants on filament length.  This
may be due to the extremely complicated numerical fitting methods used
in the treatment of the data. We have arguments, however, that the
twist and bending modes are coupled giving rise to new sources of
dissipation. Thus the friction coefficient may be underestimated in
the standard theories of twist dynamics;  the twist elastic
constant is less well characterized than the bending constant.

The theory of the twist dynamics of semi-flexible polymers has been
discussed in \cite{zimm,dna} in conjunction with DNA dynamics. The
twist dynamics of a straight filament are over-damped. Thus the twist
angel $\phi$ in the laboratory frame obeys the equation

\begin{equation}
{ {\partial \phi} \over {\partial t}} = K { {\partial^2 \phi} \over { \partial s^2}} \label{twist}
\end{equation}
with $K$ a rotational diffusion coefficient depending on the torsional
rigidity of the filament and the solvent friction coefficient.

One should note the large difference in time scales between the
relaxation of bending and torsional modes. Whereas a ten micron
filament will take several seconds to relax its fundamental bending
mode the relaxation time of its torsional modes occurs on the scale of
microseconds to milliseconds. This is due to the very different length
scales upon which dissipation takes place for bending and
translational modes. Consider a bending excitation of length scale $L$
and amplitude $a$ on a semi-flexible filament. The total energy of the
excitation is $E \sim (a^2 l_p/L^4) L$.
The dissipation in the solvent is caused by the time evolution of the
amplitude and can be estimated to be $ \eta L^3 (\dot a /L)^2 $ where
$\eta$ is the viscosity $L^3$ is the volume of fluid implicated in the
flow and $\dot a /L$ is a typical shear around the filament. Thus we
find the approximate equation in real space for the evolution of the
amplitude $a$.
\begin{equation}
\eta L \dot a \sim - a l_p /L^3
\end{equation}

or
\begin{equation}
\dot a \sim - a l_p /L^4
\end{equation}
This is the equivalent in real space of the dispersion relation
$\omega \sim q^4$ deduced above from the Langevin equation
(\ref{langevin}).

How does this argument change for twist dynamics? The hydrodynamic
field due to rotation of the filament has typical shear values of $
\dot\phi$ and occurs throughout a volume $Ld^2$. The energy stored is
$E \sim L (\phi^2 l_{r}/L^2)$. Thus we deduce the approximate equation

\begin{equation}
\eta L d^2 \dot \phi \sim - \phi l_{r} /L 
\Longrightarrow 
\dot \phi \sim - l_r \phi/d^2 L^2
\end{equation}

In agreement with (\ref{twist}) we find a $q^2$ dispersion relation
due to the short range nature of the dissipation. A filament of length
$L$ thus has two relaxation times with a ratio $\tau_{L}/\tau_{twist}
\sim (L/d)^2$. Since $L/d \sim 1000$ these times are very different
indeed. The rotational relaxation occurring of sub-millisecond time
scales is well adapted to times scales of fluorescent re-emission.
Video techniques are far to slow to capture these movements though
perhaps more sophisticated techniques such as those developed by
\cite{schmidt2} or \cite{weitz} are able to resolve such high
frequency fluctuations. Our picture of bending fluctuations should
thus be modified. During the slow relaxation of the bending modes the
filament must be considered as rotating at very high frequencies. The
bending occurs subject to a very high frequency ``noise'' due to the
rotational motion.

The de-polarizing experiments are sensitive to correlation functions
of the form
\begin{equation}
I(q,t) = \exp( - \langle (\phi(t) -\phi(t'))^2 \rangle ) = \exp( - H(t-t')).
\label{I}
\end{equation}
we can find the form of these correlation function $H(t)$ with the
following simple argument similar to that used to derive
eq.~(\ref{force}).  Eq.~\ref{twist} has the normal diffusive form thus
in a time $t$ a length $ \sim \sqrt{t} $ of filament moves coherently.
This section of length $ \sqrt{t}$ has a rotational diffusion constant
$D_r \sim 1/\sqrt{t} $ thus after a time $t$ we deduce that $\vert
\phi(t) -\phi(0) \vert \sim t^{1/4}$ (as confirmed by more elaborate
treatments). This slow decay of the rotational correlations is valid
for as long as one is dominated by the internal modes of the filament.
For longer times the filament is going to turn freely on its axis and
thus $ \vert \phi(t) -\phi(0) \vert \sim t^{1/2}$ as expected for a
diffusive process

In polydisperse samples it may well be difficult to separate the
effects of polydispersity from the internal dynamics. The signal from
a short filament has the form $I(t) \sim \exp( -t/ d^2 L )$ which when
convolved with the length distribution $ P(L) \sim \exp(-L/l_0)$ gives
a dominant contribution in $ <I(t)> \sim \exp (- (t/t_0)^{1/2})$ this
is perhaps partly the cause of the difficulty of deconvolving the
signal found in \cite{twist}.

What are the effects of bending on the rotational dynamics? Does the
filament keep a constant shape in the body frame of reference or is
the converse true, that the filament twists with a constant real space
shape, imposing a rapid bending on the filament as it spins. The first
scenario is rather unlikely due to the high frictional coefficient of
a bent object turning in space. Since the typical transverse
fluctuations of a filament of length $L$ vary as $L^{3/2}/l_p^{1/2}$
one can redo the above arguments as to the typical gradients and
volumes involved in dissipation to find that the rotational diffusion
coefficient should vary as $D_r \sim l_p/L^4$ rather than the more
usual $D_r \sim 1/d^2 L$. It is more reasonable to expect that the
approximate torsional mode is a {\it combined } \/ spin and bend such
that the shape of the filament remains the same in the laboratory
reference frame. Indeed one can formally let the bending friction
diverge in which case this combined bending and torsional mode becomes
the exact dynamical eigenmode of the filament.  Clearly, as well, in
dense solutions this must be the physical torsional mode due to the
strong steric hindrance present. Note that due to the non-diagonal
nature of the frictional couplings the eigenmodes for the {\it statics
  }\/ and for the {\it dynamics}\/ can be expected to be very
different.

Since in the frame of the rotating filament, the filament is bending
very rapidly we can find new sources of dissipation in the dynamics.
Until know we have considered that the filament is in effect a {\it
  perfect solid}\/ with no dissipation occurring due to internal
motion, however, in general a protein is expected to be described by
both an elastic modulus, $E$ and a loss modulus. The simplest model is
\begin{equation}
E(\omega) = E/(1 + i \omega \tau_i)
\label{GGG}
\end{equation}
Where $\tau_i$ is an internal frequency that one could perhaps place
at about 1MHz. Thus by avoiding spinning in the laboratory frame the
one finds a new source of dissipation, internal friction. It is to be
seen whether these effects are large enough to be seen experimentally.




As well as coupling to twisting modes, the bending modes also
contribute to the depolarized scattering; bending a filament changes
its orientation in space giving a contribution to the rotation of the
filament analogous to (\ref{GG})
  
\begin{eqnarray}
\label{final}
<(\phi(t) -\phi(0))^2> & \sim &
\int q^2 { {1- \exp(-2 \kappa q^4 t/\rho)} \over {\kappa q^4} } {{dq}\over{2 \pi} }\\
& =  & B t^{1/4}/\kappa^{3/4}
\end{eqnarray}
Since the twist modes are so much more rapid than the bending modes we
would normally expect that this contribution is to slow to be
observed, however in the presence of crosslinking agents which freeze
some of the twist degrees of freedom one might hope to observe such
terms.

\begin{figure}
  \epsfxsize=150pt \centerline{\epsfbox{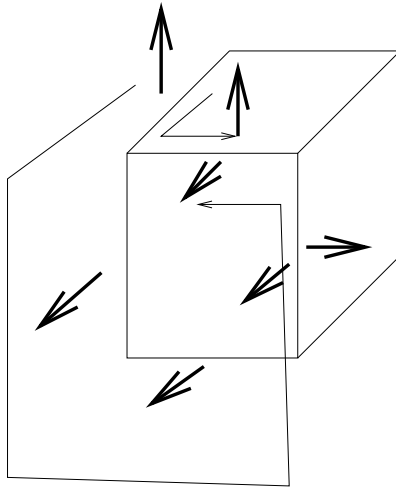}}
\caption[path]{
\label{path}
Illustration of the relative rotation of a filament due to the non
commutative nature of the rotation group in three dimensions. We start
at the vertical arrow on the upper side of the cube and follow two
paths. The first goes clockwise around the from face, the second
anti-clockwise. At each corner we imagine that we are allowed to bend
a filament but not twist it. We see that an the arrows finish at 90
degrees to one another. Thus we can twist the filament in space by
only bending distortions. Bending fluctuations of a filament can thus
give arise to an apparent twist of the filament in the laboratory
frame, which adds to the internal twist dynamics of the filament. }
\end{figure}

It will be interesting to think about the problem of back reaction of
the rotational degrees of freedom on the bending.  At the shortest
length and time scales the bending and twist should be weakly coupled;
whereas as we have argued above in long filaments the twisting
eigenmode is really a combined twist/bend mode. Fig. (\ref{path})
shows that the bending and twisting degrees of freedom are indeed
strongly coupled in long filaments. Is there a dynamic crossover
between a short time regime where the bending and the twisting of a
filament do not see each other and a long time regime where the modes
are coupled dynamically?  It may be possible to make an adiabatic
expansion in the dynamic equations using the enormous difference in
time scales as a separation parameter to answer this question.  Such
effects may give crossovers in the dynamic light scattering
experiments especially at shorter times when the ratio of
characteristic time for twist and bend are not too different.

A simple simulation, \cite{me2}, to measure the coupling between the
twist and bend described by fig. (\ref{path}) shows that the filament
twists in the laboratory frame in the following manner: $\vert \phi(s)
- \phi(0) \vert \sim s/l_p$. Thus from a simple scaling argument we
expect that the mechanism of fig. (\ref{path}) contributes a term
$\vert \phi(t)-\phi(0) \vert \sim t^{1/4}/\kappa^{3/4}$ to the
scattering (\ref{I})

Ref. \cite{siggia} has shown that small asymmetries in coupling
constants can give rise to a {\it static}\/, though weak, cross talk
between torsion and bending. Similar effects are also probable in the
case of actin filament though are probably as in DNA small
corrections. Here we note that twist-bend coupling terms are always
present in the dynamics due to the non-diagonal nature of the
hydrodynamic couplings, even when the modes are statically
independent.

\subsection*{Actin forms active materials}

Reversible and irreversible crosslinking proteins exist with several
attachment geometries. Again to many such proteins exist to be fully
treated here \cite{molecules} but one can note in particular the
existence of proteins with variable crosslinking properties. The
exact geometry of interaction varies with a tendency to form parallel
bundles for the case of $\alpha$-actinins and perpendicular link for
actin binding protein-120. The crosslinks can be either irreversible
or reversible depending upon the molecule involved and the exact
chemical nature of the buffer (many molecules are sensitive to energy
sources such as ATP, and signaling ions such as calcium).

Until know we have discussed actin in the relation to classical
mechanical and rheological properties. However, actin is a molecule
which crucial in the transduction of chemical into mechanical energy
and thus is potentially much richer than classical polymer systems.
Already the filament undergoes a active polar polymerization process
which means that a single filament can move forwards by depolymerizing
behind while polymerizing in front, a process known as tread-milling
\cite{treadmilling}.

In the cell actin has many independent roles. It is a major component
of muscle where it is present in the form of aligned fibers. However
it is also present in many other cells in both higher and lower
eucaryotes (such as yeast or slime molds). In these cells the actin is
often present in the form of a complex gel like environment forming
the cellular cortex. This structure is essential to movement of many
cells.

In both muscle and in other cells the actin is associated with active
molecules, in particular molecules from the {\it myosin}\/ family
\cite{block}. These molecules are molecular motors, that is they
transform chemical energy (in the form of ATP) into mechanical force.
Recent experiments extract myosin from the cell and use it to coat
either surfaces or beads. The filaments then slide over the glass
surface or the beads are transported along an actin filament as a
consequence. It would be of the greatest interest to study the effect
of myosin in a gel. One could imagine a solution with multiple beads
where the motors on each bead interact with several filaments. These
experiments appear quite difficult to perform, however recent results
similar to this are available for the more rigid microtubule in the
presence of a kinesin construct\cite{nedelec}.

One of the most remarkable manifestations of the variety of behaviors
shown by actin is the listeria locomotion system. Listeria is a
pathogenic bacterium of the gut which is able to catalyst the
polymerization of free cellular actin into hollow tube. By an unknown
mechanism the bacterium can ``push'' against the tube to propulse
itself forwards. Highly simplified extracts now exist where the
bacterium remains active; the system in being actively studied by a
number of biophysics groups, in particular to characterize the
mechanical properties of this actin tube, made out of short
crosslinked filaments.

\vskip 10pt 

\small The author would like to acknowledges the help of F. Amblard,
M-F. Carlier, E. Farge, H. Isambert, P. Janmey J. Kas S. Leibler F.
MacKintosh and E. Sackmann for the many discussion involved in this
work


\begin{thebibliography}{99}
\bibitem{protocol} A part of the polydispersity in actin solutions comes from the fact that growth is a nucleated process starting at different moments in the sample. It we have a seed in the form of gelsolin and block the dynamics of depolymerization with phalloidin the final solution should be relatively mono-disperse.
\end{thebibliography}
\end{document}